\documentclass[aps,prl,twocolumn,amsmath,amssymbs,superscriptaddress,longbibliography]{revtex4-1}
\usepackage{epsfig}
\usepackage{wrapfig}
\usepackage{bbm}
\usepackage[usenames]{color}
\usepackage{array}
\usepackage{times}

\usepackage{graphicx}
\usepackage{float}
\usepackage{multirow}
\usepackage{natbib,twoopt}
\usepackage{xcolor}
\usepackage{amsmath,amsfonts,amssymb}
\usepackage{graphicx}
\usepackage{color}
\usepackage{stackengine}
\usepackage{subfigure}
\usepackage{verbatim}
\usepackage{stmaryrd} 


\usepackage{soul}

\newcommand{\RNum}[1]{\uppercase\expandafter{\romannumeral #1\relax}}

\newcommand{\balancecolsandclearpage}{%
	\close@column@grid
	\cleardoublepage
	\twocolumngrid
}

\begin{document}

\title{Photonic molecule approach to multi-orbital topology}


\author{Maxim Mazanov}
\affiliation{School of Physics and Engineering, ITMO University, Saint  Petersburg 197101, Russia}

\author{Diego Rom\'an-Cort\'es}
\affiliation{Departamento de Física and Millenium Institute for Research in Optics–MIRO, Facultad de Ciencias Físicas y Matemáticas, Universidad de Chile, 8370448 Santiago, Chile}

\author{Gabriel Cáceres-Aravena}
\affiliation{Departamento de Física and Millenium Institute for Research in Optics–MIRO, Facultad de Ciencias Físicas y Matemáticas, Universidad de Chile, 8370448 Santiago, Chile}
\affiliation{Institute of Physics, University of Rostock, 18051 Rostock, Germany}


\author{Christofer Cid}
\affiliation{Departamento de Física and Millenium Institute for Research in Optics–MIRO, Facultad de Ciencias Físicas y Matemáticas, Universidad de Chile, 8370448 Santiago, Chile}

\author{Maxim A. Gorlach}
\affiliation{School of Physics and Engineering, ITMO University, Saint  Petersburg 197101, Russia}

\author{Rodrigo A. Vicencio}
\affiliation{Departamento de Física and Millenium Institute for Research in Optics–MIRO, Facultad de Ciencias Físicas y Matemáticas, Universidad de Chile, 8370448 Santiago, Chile}
\email{rvicencio@uchile.cl}

\begin{abstract}
 The concepts of topology provide a powerful tool to tailor the propagation and localization of light. While electromagnetic waves have only two polarization states, engineered degeneracies of photonic modes provide novel opportunities resembling orbital or spin degrees of freedom in condensed matter. Here, we tailor such degeneracies for the array of femtosecond laser written waveguides in the optical range exploiting the idea of photonic molecules~-- clusters of strongly coupled waveguides. In our experiments, we observe the emergence of topological modes caused by the inter-orbital coupling and track multiple topological transitions in the system with the change of the lattice spacings and excitation wavelength. This strategy opens an avenue in designing novel types of photonic topological phases and states.
\end{abstract}

\maketitle

\section{Introduction}
Topological photonics provides a versatile approach to shape the light flows by harnessing the topology of photonic bands~\cite{Lu2014Nov,Lu2016,Ozawa_RMP_2019}. This can be achieved by the variety of ways including plasmonic~\cite{Cheng2015}, photonic~\cite{Rechtsman2013,Hafezi2013,Barik2018Feb} or polaritonic~\cite{Karzig2015,Klembt2018} systems operating from radiofrequency~\cite{Yves2017,Yang2019} to the optical~\cite{Rechtsman2013,Noh2018} domain. 

Experimental realization of optical topological structures is especially demanding because of relatively narrow topological bandgaps and losses. The major strategies to create topological structures for the optical domain include arrays of coupled ring resonators with engineered couplings~\cite{Hafezi2011,Hafezi2013,Mittal_2019} or crystalline topological structures~\cite{Wu_Hu,Barik2018Feb,Noh2018,Gorlach2018,Para2020}. While the former approach covers mostly the infrared range being hard to scale towards visible wavelengths, the latter strategy requires carefully designed lattice symmetry and does not provide much flexibility in reconfiguring the topological properties, creating pseudo-spin degrees of freedom or tailoring effective spin-orbit interactions. Therefore, these approaches do not unlock full functionalities available for photonic topological structures. 

A recently suggested alternative is to optimize the responses of the {\it individual meta-atoms} achieving an accidental degeneracy of their modes~\cite{Caceres-Aravena2019Jul,Aravena_PRA_2020,Silva_PRL_2021,Caceres-Aravena2022Jun, 
SavelevGorlach_PRB, 
Mazanov2022May,Mikhin2023, 
Huang_2011, Sakoda_2012_3, Gorbach:23}.
Such degenerate modes play a role analogous to the orbital or spin degrees of freedom in condensed matter or cold atom optical lattices~\cite{Li_2016, Li_2013, Pelegri_PRA_2019, Yin_superfluid_2015,Wu_PRL_2007}. Since the number and the symmetry of the degenerate modes can be tailored on demand, this opens up rich plethora of novel topological phases. While this appealing strategy can be implemented directly at microwave frequencies~\cite{Mazanov2022}, its realization in the visible range is severely hindered due to the fabrication limitations.



Here, we resolve this conceptual difficulty by experimentally realizing the concept of \textit{photonic molecules}~-- a set of closely spaced waveguides which together give rise to photonic orbitals with different symmetries. Employing femtosecond laser writing technique~\cite{Davis96,Szameit2005,Flamini2015}
~-- an approach that has consolidated as a key technique in physical lattice studies~\cite{floquet,Liebus,Liebseba,corner2019}~--
to control the refractive index contrast, we create  multi-orbital photonic lattices such that fundamental $s$ (symmetric or bonding) and horizontal excited $p$ (antisymmetric or antibonding) modes become degenerate and interact with each other enabling a richer physics. Specifically, we experimentally track topological transitions in our system upon change of the wavelength and lattice spacings observing multiple topological states.

\begin{figure*}[t!]
	\centering
	\includegraphics[width=0.99\textwidth]{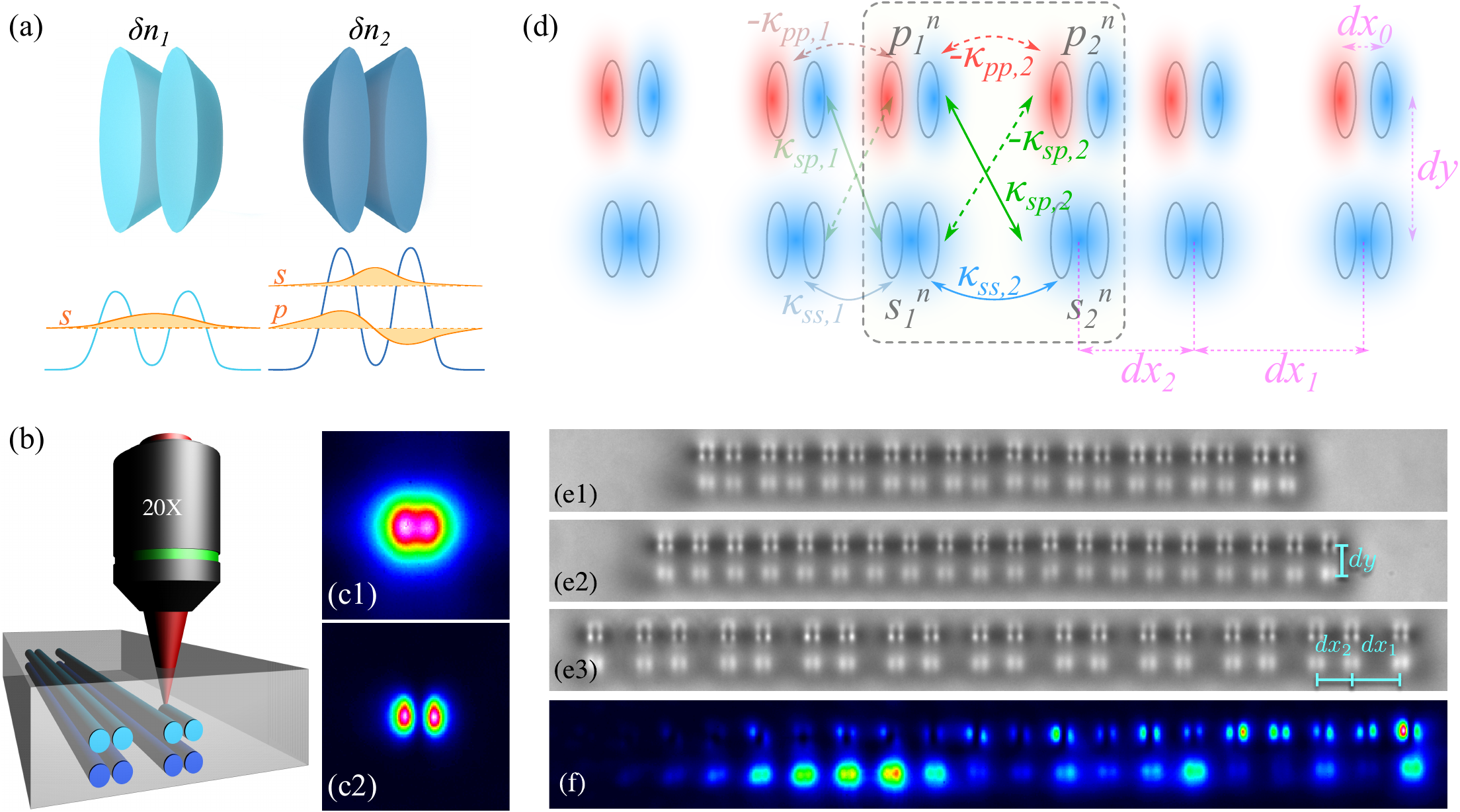}
	\caption{ 
        \textbf{Photonic molecules and the $s$-$p$ hybridized Su-Schrieffer-Heeger lattice.} 
        (a) Photonic molecules concept. Top: Each molecule is composed of two identical waveguides. We engineer the degeneracy of $s$ and $p$  modes in two adjacent molecules using different refractive index contrast. 
        (b) Illustration of the femtosecond laser writing technique, where light blue (blue) color represents the $s$ ($p$) molecules obtained for the different fabrication power. (c1) $s$ and (c2) $p$ mode intensity profiles obtained after exciting the respective photonic molecule with a $730$ nm laser beam. 
        (d) Sketch of the multi-orbital Su-Schrieffer-Heeger-type lattice. Photonic molecules support near-degenerate $s$ or $p$ modes. Arrows designate the nearest-neighbour couplings, the dashed box shows the unit cell. 
        (e) White light microscope images of three fabricated lattices with the lattice spacing $dx_1:$ (e1) $19$~$\mu$m, (e2) $25$~$\mu$m and (e3) $35$~$\mu$m, for $dx_2=25$~$\mu$m and $dy=20$~$\mu$m. (f) Intensity output profile after exciting the $s$ molecule at the right edge of the homogeneous lattice shown in (e2).}  
	\label{fig:F1}
\end{figure*}


%
%
\twocolumngrid

\section{Results}

{\bf Photonic molecules.}~\label{sec:1} The simplest photonic molecule consists of a couple of closely spaced identical elliptical waveguides [see Fig.~\ref{fig:F1}{\bf a}-top], which give rise to the pair of symmetric ($s$) and anti-symmetric ($p$) horizontal modes with the typical intensity profiles shown in Fig.~\ref{fig:F1}{\bf c}. While the propagation constants of $s$ and $p$ modes in a given molecule are strongly different, they can be matched for the adjacent molecules by tuning the refractive index contrast of the waveguides~\cite{Silva_PRL_2021,Caceres-Aravena2022Jun} [see Fig.~\ref{fig:F1}{\bf a}-bottom] during the laser writing process illustrated schematically in Fig.~\ref{fig:F1}{\bf b}.

Although the molecular idea is straightforward, its experimental implementation is nontrivial and faces several challenges. First, due to the ellipticity of the waveguide geometry, vertical $p$ modes strongly affect the behavior of a single waveguide. However, if a molecule is assembled properly, $s$ and horizontal $p$ modes dominate. Another issue is the choice of the distance $dx_0$ between the waveguides comprising the molecule [see Fig.~\ref{fig:F1}{\bf d}]. From one side, this distance should be larger than the typical width of a single waveguide, which is around $ 4$~$\mu$m~\cite{Szameit2005}.
On the other hand, larger spacings in between the waveguides deteriorate their effective coupling and prevent the formation of clear $s$ and $p$ molecular modes. After extensive experimental investigations, we optimized the distance $dx_0=7\ \mu$m, which yields clear $s$ and $p$ horizontally disposed degenerate modes, as the ones shown in Fig.~\ref{fig:F1}{\bf c}. This trade-off depends strongly on specific fabrication parameters, such as the writing power and the writing velocity, as well as the wavelength of excitation (see Supplementary Materials).

In femtosecond laser-written lattices~\cite{Szameit2005}, a standard center-to-center distance of $16\ \mu$m is taken as a lower limit for which coupled mode theory (tight-binding approach) is still valid. Thus, due to the relatively small distance between the waveguides in our molecular setup, the tight-binding model does not describe the formation of molecular $s$ and $p$ orbitals. However, the interaction of the individual molecules within the molecular lattice is well captured by the tight-binding approach, which allows us to develop a consistent theoretical description.

\begin{figure*}[t!]
	\centering
	\includegraphics[width=0.99\textwidth]{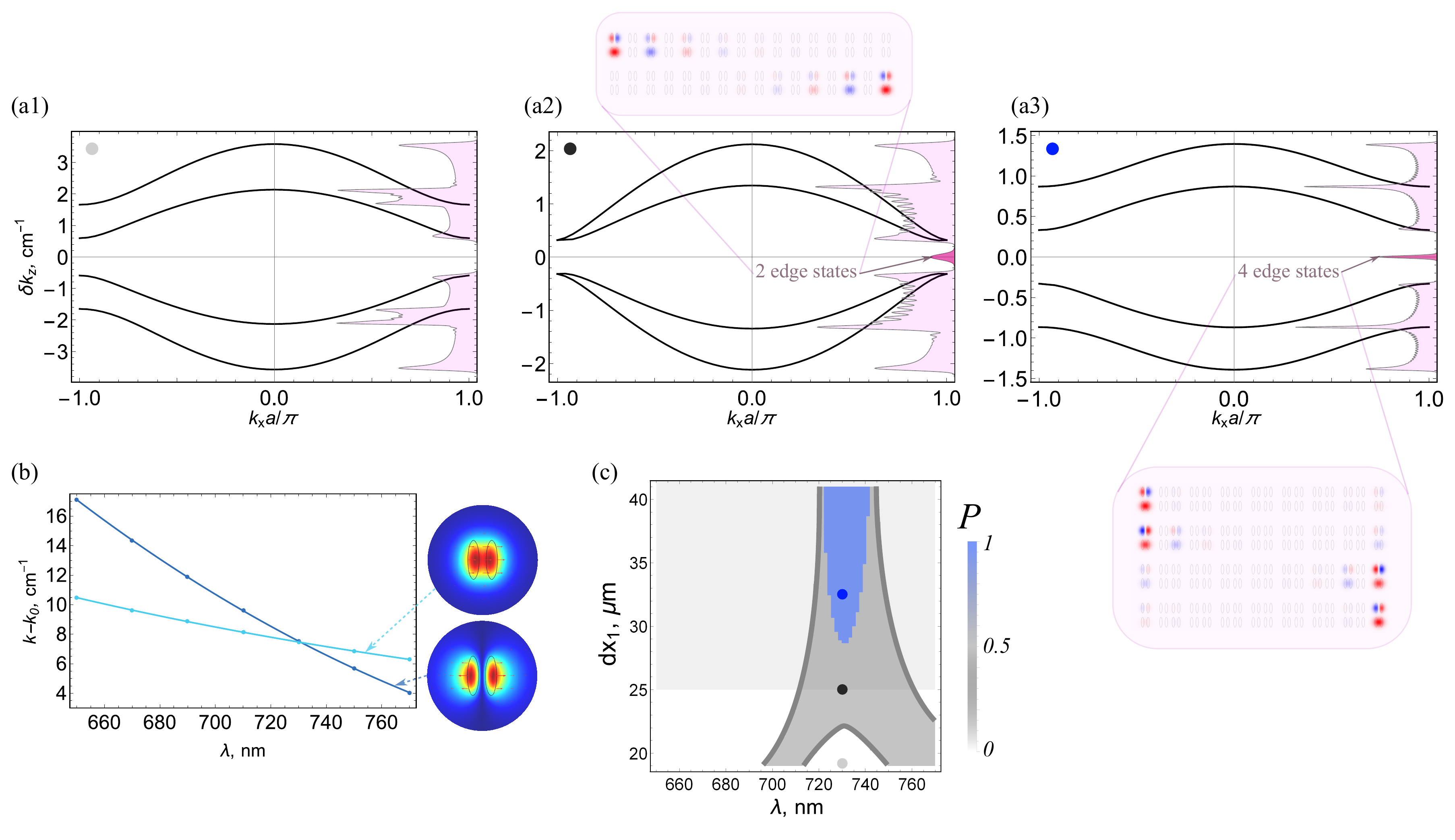}
	\caption{ 
        \textbf{Band topology and edge states for the molecular SSH lattice.} (a)~Spectra of propagation constants calculated from Eq.~\eqref{Hamk} for $dx_1=19$~$\mu$m (a1), $25$~$\mu$m (a2) (corresponding to the models discussed in Refs.~\cite{Aravena_PRA_2020, SavelevGorlach_PRB}) and $33$~$\mu$m (a3), for
        $dx_2=25$~$\mu$m and $dy=20$~$\mu$m. Insets on the right show the corresponding density of states for the finite $30$-cell lattice, 
        with hybrid-orbital edge states contribution highlighted in dark purple. Graphs near the panels (a2-3) show the calculated edge state profiles in the tight-binding model of an $8$-cell lattice. 
        (b)~Numerically computed dispersion of the $s$ and $p$ modes indicating perfect degeneracy at $730$~nm (parameters: $\delta n_1 = 4.132 \cdot 10^{-4}$, $\delta n_2 = 8.836 \cdot 10^{-4}$). 
        (c)~Topological phase diagram for the the Wannier center polarization $P$ of the central bandgap calculated numerically from the tight-binding model. The region shaded by light gray marks the nontrivial Zak phase of the first (lowest) and third (highest) bandgaps indicating the appearance of practically single-mode edge states.  
        }
	\label{fig:F2}
\end{figure*}

{\bf Su-Schrieffer-Heeger model with inter-orbital coupling.} As a simple yet nontrivial example of photonic molecules paradigm, we design and implement a one-dimensional photonic waveguide lattice shown schematically in Fig.~\ref{fig:F1}{\bf d}. Note that the molecules in the upper and lower rows have different refractive index contrast $\delta n$ which allows us to match the propagation constant of $s$ modes in the lower row to that of the $p$ modes in the upper row, and thus enabling multi-orbital physics. The molecules comprising a single column do not interact with each other due to symmetry reasons.

Since the distance in between the adjacent molecules experimentally exceeds $19$~$\mu$m, the physics of the system is captured by the tight-binding model. In the case of equidistant molecular lattice, this requires three positive coupling parameters: $\kappa_{ss}$, $\kappa_{pp}$ and $\kappa_{ps}=-\kappa_{sp}$, describing the interaction of $s$ and $s$, $p$ and $p$, $s$ and $p$ orbitals, respectively. The signs of the coupling constants are defined by the overlap of the respective mode fields~\cite{Aravena_PRA_2020,SavelevGorlach_PRB}.

To further exploit the versatility of our platform, we fabricate a set of experimental structures introducing the dimerization in the lattice spacings ($dx_1$,$dx_2$) between the molecules similarly to the well-celebrated Su-Schrieffer-Heeger model~\cite{1DSSH}, as described in Figs.~\ref{fig:F1}{\bf d} and {\bf e}. As a result, the physics of our system is defined by the interplay of the two factors: interference of nearly degenerate $s$ and $p$ molecular orbitals and geometric dimerization mechanism.

Developing the theoretical description of our system, we adopt the basis of molecular modes $\textbf{v}^{n}=(s_1^n,p_1^n, s_2^n,p_2^n)^T$, with the unit cell of the lattice shown in Fig.~\ref{fig:F1}{\bf d}. In these designations, the tight-binding Bloch Hamiltonian reads:
%
\begingroup
\begin{equation}
\label{Hamk}
\hat{H}(k_x)
=
\begin{pmatrix}
	\hat{\beta} & e^{-i k_x} \hat{\kappa}_{1-} + \hat{\kappa}_{2+} \\
        e^{i k_x} \hat{\kappa}_{1+} + \hat{\kappa}_{2-} & \hat{\beta} \\
\end{pmatrix}
,
\end{equation}
\endgroup
where $ k_x$ is the Bloch wave number, $ \hat{\beta} = \frac{1}{2}\,\delta k_{z0} \cdot \hat{\sigma}_z$ and $\delta k_{z0}$ is the difference in the propagation constants of $s$ and $p$ orbital modes in an isolated photonic molecule. 
The coupling matrices read: $ \hat{\kappa}_{i \pm} = \left(
\begin{array}{cccccc}
-\kappa_{ss,i} & \pm \kappa_{sp,i} \\
\mp \kappa_{sp,i} & \kappa_{pp,i} \\
\end{array}
\right) $, with $\kappa_{ss,i}$, $\kappa_{sp,i}$ and $\kappa_{pp,i}$ being the positive inter-mode coupling constants that depend on the distances $dx_{1,2}$ and $dy$ between the molecules, as described in Fig.~\ref{fig:F1}{\bf d}. For quantitative predictions, we estimate these coupling coefficients numerically from the splitting of the modes of the two interacting photonic molecules (see Supplementary Materials). The eigenvalue equation then takes the form 
\begin{equation}\label{eq:Eigenvalue}
\hat{H}(\textbf{k}) \textbf{v} = (k_z-k_0) \textbf{v}\:,
\end{equation}
where $k_z$ is the propagation constant of a collective mode in a waveguide array and $k_0$ is the half-sum of the propagation constants of $s$ and $p$ modes in an isolated molecule.


{\bf Theoretical model.} While the model above looks simple from the first glance, it features a plethora of topological phases. From an experimental perspective, this  physics can be mapped by changing two tuning parameters. One of them is the lattice spacings $dx_1$ and $dx_2$, whose ratio defines the dimerization of the lattice and varies from sample to sample. Another one is the excitation wavelength which controls the detuning between the propagation constants of $s$ and $p$ molecular modes, with a degeneracy condition achieved at $\lambda\approx 730$~nm in the experiments.

By solving the eigenvalue problem Eq.~\eqref{eq:Eigenvalue} with the Hamiltonian Eq.~\eqref{Hamk}, we recover the spectra of the propagation constants in Fig.~\ref{fig:F2}{\bf a}. Interestingly, all calculated spectra are symmetric with respect to the ``zero level'' corresponding to the propagation constant equal to $k_0$. This feature reflects chiral symmetry of the model at perfect mode degeneracy $\delta k_{z0} = 0$ captured by the chiral symmetry operator 
\begin{equation*}
    \hat{\Gamma}=
    \begin{pmatrix}
        \hat{I} & 0\\
        0 & -\hat{I}
    \end{pmatrix}\:.
\end{equation*}

If the lattice is terminated by a strong coupling link, i.e. the lattice spacing $dx_1$ at the edge is the smallest, the structure does not support any edge states [Fig.~\ref{fig:F2}{\bf a1}]. In other words, dimerization wins the competition with the mechanism of the mode interference preventing the formation of the edge-localized states.

However, once the lattice is made equidistant, the interference between $s$ and $p$ modes becomes increasingly important and results in the formation of a pair of hybrid edge-localized modes [Fig.~\ref{fig:F2}{\bf a2}] in line with the earlier theoretical works~\cite{Aravena_PRA_2020,SavelevGorlach_PRB}. The pairwise degeneracy of the bands at the edges of the Brillouin zone [Fig.~\ref{fig:F2}{\bf a2}] indicates the possibility to choose a smaller unit cell with just two photonic molecules thus unfolding the Bloch bands. Finally, terminating the lattice by a weak coupling link ($dx_1>dx_2$), we ensure that the effects of the lattice dimerization and mode interference add up, which results in the formation of four edge-localized modes existing in a single bandgap [Fig.~\ref{fig:F2}{\bf a3}].



Since bulk polarization in inversion-symmetric systems can take only two values, 0 and $1/2$, it appears insufficient to capture the scenarios Fig.~\ref{fig:F2}{\bf a1}-{\bf a3} of the edge mode formation. This difficulty can be resolved by interpreting $(s+p)$ and $(s-p)$ mode combinations as effective spin degrees of freedom, retrieving the bulk polarization for each of the pseudospin blocks separately and adding up the contributions. In the general case of nonzero coupling between the introduced pseudospins, the topological invariant is computed as the expectation value of the position operator projected on the two lowest energy bands (see Supplementary Materials).

Retrieving the polarization this way using the numerically obtained couplings and mode detuning (see Fig.~\ref{fig:F2}{\bf b}), we recover the phase diagram presented in Fig.~\ref{fig:F2}{\bf c}. 
In the standard Su-Schrieffer-Heeger model, we expect zero bulk polarization for $dx_1<dx_2$ ($dx_2=25$~$\mu$m) and nonzero result for $dx_1>dx_2$. 

However, the topological phase diagram of our system appears to be richer as shown in Fig.~\ref{fig:F2}{\bf c}. If $s$ and $p$ modes are perfectly degenerate, the central bandgap hosts 0, 2 or 4 edge-localized modes, consistent with the calculated polarization via the bulk-boundary correspondence. This situation persists for moderate detunings between the propagation constants of $s$ and $p$ modes (see the region delimited by the gray thick line). However, once the wavelength is shifted further away from the optimum value of $730$~nm, the edge modes in the central bandgap disappear. In this parameter range, the mechanism of mode interference is no longer effective, and hence the spectra for $s$ and $p$ modes decouple. As a consequence, $s$ and $p$ modes develop their own, practically independent, spectra and topological edge states appear under the usual condition $dx_1>dx_2$. However, they are no longer hybrid and exist outside of the central bandgap.

\begin{figure*}[t!]
	\centering
	\includegraphics[width=0.99\textwidth]{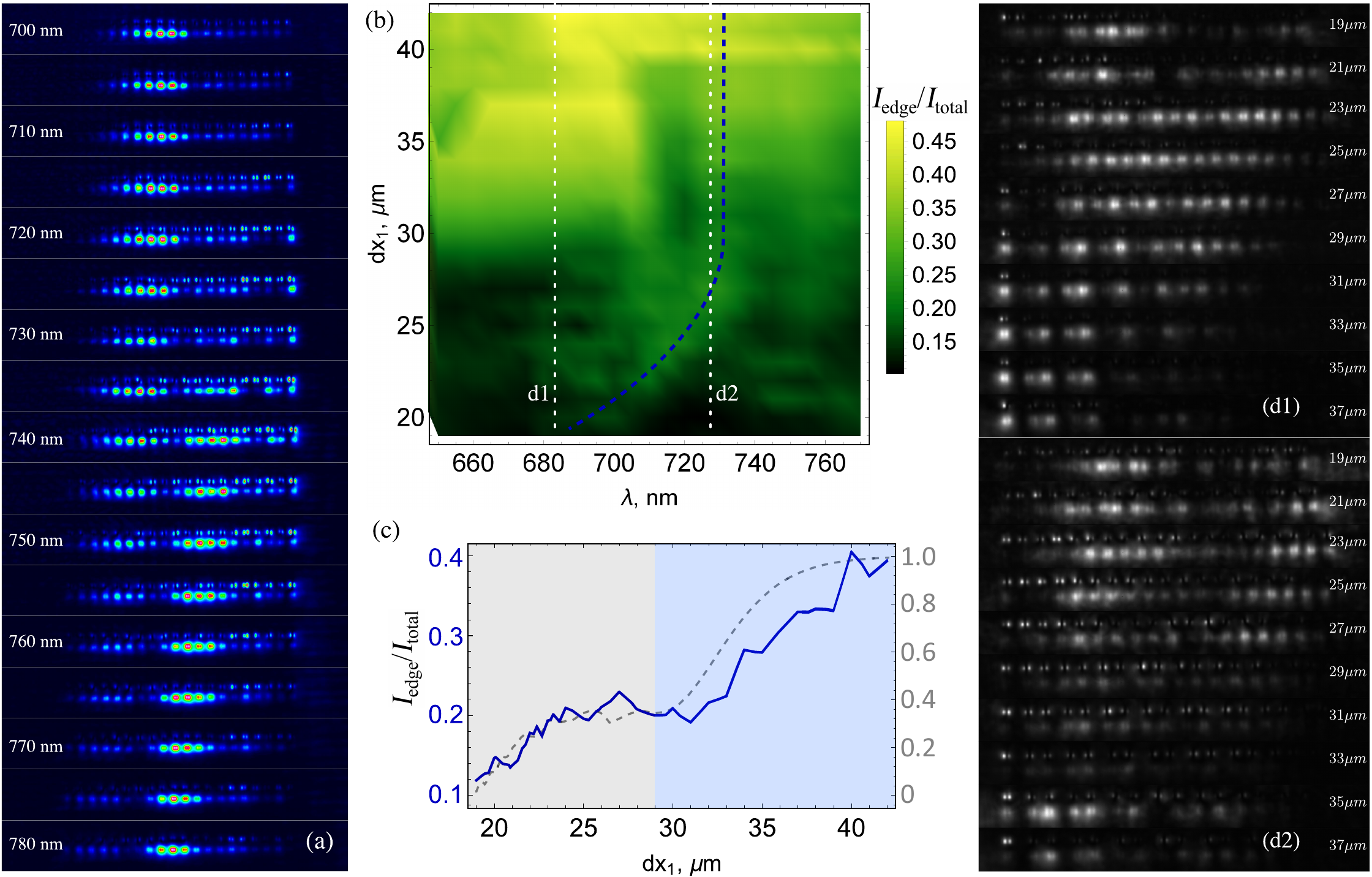}
	\caption{ 
        %
        \textbf{Experimental studies of topological lattices.} (a) Intensity output profiles after exciting the homogeneous lattice ($dx_1=dx_2=25\ \mu$m) at the lower right 
        edge $s$ molecule at different wavelengths $\lambda=\left(700,780\right)$~nm.
        (b) Experimental edge intensity distributions averaged with the first $3$ edge columns, for left and right $s$ mode edge excitation versus $\lambda$ and $dx_1$, for fixed $dx_2 = 25$ and $dy=20\ \mu$m. 
        (c) Comparison of experimental (blue) and re-scaled numerical (dotted gray) edge intensity distributions along the dotted blue line shown in (b). A topological transition between the two phases highlighted by gray and blue is observed.
        (d) Intensity output profiles for the lattice spacings $dx_1$ indicated at the figure, after exciting the lattice at the left edge $s$ molecule for an excitation wavelength (d1) $680$~nm and (d2) $725$~nm, corresponding to the vertical white dashed lines in panel (b).
        }
	\label{fig:F3}
\end{figure*}

{\bf Experiments.} To probe our predictions experimentally, we fabricate 25 photonic molecular lattices, having a 70~$\text{mm}$ propagation length, with fixed $dy = 20 \mu \text{m}$, $dx_2 = 25 \mu \text{m}$ and $dx_1$ varying from $18 \mu \text{m}$ to $42\mu \text{m}$ (see Supplementary Materials). 
We excite those structures via the lower ($s$) photonic molecule either from the left or from the right edge of the array and record the output intensity distribution at different excitation wavelengths, by using a supercontinuum laser source. 

First, we examine the results for an equidistant molecular lattice [see Fig.~\ref{fig:F3}{\bf a}] in the wavelength range from 700~nm to 780~nm, every $5$ nm. Due to the chosen excitation scheme, the output intensity of the $s$ sublattice is generally larger than the $p$ one. 
However, those intensities become comparable at wavelengths around $720-750$~nm indicating an efficient inter-orbital coupling. In the same wavelength range, the output field exhibits the localization at the right edge of the array. This points towards the formation of topological edge states possessing a hybrid origin; i.e., having the contribution both from $s$ and $p$ molecular modes.

To further visualize the emergence of topological edge states in our system, we plot the intensity at the edges of the structure as a function of the excitation wavelength $\lambda$ and the lattice spacing $dx_1$ [Fig.~\ref{fig:F3}{\bf b}]. Here, brighter areas correspond to the edge-localized modes, while darker spots show the lack of edge localization arising in the trivial regime and in  topological transition points. However, the diagram in Fig.~\ref{fig:F3}{\bf b} does not allow to differentiate the topological phases with 2 and 4 edge states. To assess this, we choose a specific trajectory on this diagram shown by a dashed blue line. If the parameters of the system are varied along this trajectory, 
analytical calculations suggest that the normalized edge intensity reaches a plateau during the topological transition. Experimental results agree well with this analytical expectation [see Fig.~\ref{fig:F3}{\bf c}] allowing to approximately capture the topological transition point.

Finally, we examine the role of the lattice dimerization for two representative wavelengths. At wavelength $\lambda=680$~nm, $s$ and $p$ modes are not degenerated, and hence the formation of the edge modes is defined by the condition $dx_1>dx_2$. In agreement with this, we observe the formation of the edge-localized field distribution for the lattice spacings $dx_1$ above 27~$\mu$m [Fig.~\ref{fig:F3}{\bf d1}]. In this regime, $s$ mode provides the dominant contribution to the edge intensity. However, the situation changes when the wavelength is shifted towards $725$~nm [Fig.~\ref{fig:F3}{\bf d2}] corresponding to the efficient $s-p$ coupling (close to zero detuning). The modes of the lattice develop a hybrid structure, while some localization is seen even when the lattice spacing $dx_1$ is slightly lower than $dx_2$. Hence, the variations of the two control parameters~-- wavelength and lattice spacing~-- enable an exquisite control over topological phases existing in the system.

\section{Discussion and conclusions}\label{sec:Discussion}

In summary, the conventional approaches to design topological phases often rely on engineering of the lattice structure. However, manipulating the response of the individual meta-atoms by tailoring the degeneracies of molecular modes with desired symmetry appears to be a promising alternative. Here, we bring this strategy to the realm of optical nanostructures, revealing the interplay of inter-orbital coupling with geometric mechanisms and imaging arising topological states experimentally.

As a consequence of those mechanisms, our system allows four topological states to coexist in the same bandgap~-- a feature, previously found only in the lattices with a special pattern of long-range interactions~\cite{Maffei2018Jan} and absent in feasible superlattice proposals~\cite{Longhi2019May,Midya2018Oct,Wei2023Jul_2nrootrealization}. In turn, this physics can be further harnessed to coherently control the topological states.

We anticipate that our approach unlocks an entire plethora of topological phenomena including higher-order topological physics and flat bands at optical wavelengths with possible generalizations towards other platforms such as polaritonics or acoustics.

\section*{Acknowledgments}

Theoretical models were supported by Priority 2030 Federal Academic Leadership Program. 
Experimental studies were supported by Millennium Science Initiative Program ICN17$\_$012, and FONDECYT Grants 1191205 and 1231313. B.R. acknowledges ANID FONDECYT de postdoctorado No.~3230139. M.M. and M.A.G. acknowledge partial support from the Foundation for the Advancement of Theoretical Physics and Mathematics ``Basis''. G.C.-A. acknowledges the IRTG  ``Imaging Quantum Systems" No.~437567992 by the German Research Foundation.

\bibliography{refs}





\end{document}


\title{Supplemental Materials: Photonic molecule approach to multi-orbital topology}

\author{Maxim Mazanov}
\affiliation{School of Physics and Engineering, ITMO University, Saint  Petersburg 197101, Russia}

\author{Diego Rom\'an-Cort\'es}
\affiliation{Departamento de Física and Millenium Institute for Research in Optics–MIRO, Facultad de Ciencias Físicas y Matemáticas, Universidad de Chile, 8370448 Santiago, Chile}

\author{Gabriel Cáceres-Aravena}
\affiliation{Departamento de Física and Millenium Institute for Research in Optics–MIRO, Facultad de Ciencias Físicas y Matemáticas, Universidad de Chile, 8370448 Santiago, Chile}
\affiliation{Institute of Physics, University of Rostock, 18051 Rostock, Germany}


\author{Christofer Cid}
\affiliation{Departamento de Física and Millenium Institute for Research in Optics–MIRO, Facultad de Ciencias Físicas y Matemáticas, Universidad de Chile, 8370448 Santiago, Chile}

\author{Maxim A. Gorlach}
\affiliation{School of Physics and Engineering, ITMO University, Saint  Petersburg 197101, Russia}

\author{Rodrigo A. Vicencio}
\affiliation{Departamento de Física and Millenium Institute for Research in Optics–MIRO, Facultad de Ciencias Físicas y Matemáticas, Universidad de Chile, 8370448 Santiago, Chile}
\email{rvicencio@uchile.cl}

\maketitle

\onecolumngrid

\setcounter{equation}{0}
\setcounter{figure}{0}
\setcounter{table}{0}
\setcounter{page}{1}
\setcounter{section}{0}
\makeatletter
\renewcommand{\theequation}{S\arabic{equation}}
\renewcommand{\thefigure}{S\arabic{figure}}
\renewcommand{\bibnumfmt}[1]{[S#1]}
\renewcommand{\citenumfont}[1]{S#1}

\tableofcontents

\section{I. Chiral symmetry}

If $s$ and $p$ modes are degenerate, i.e. $\delta k_{z0}=0$, the Hamiltonian of our model, Eq.~(1) of the main text features an internal symmetry known as chiral symmetry. Specifically, the operator
\begin{equation}
    \hat{\Gamma}=
    \begin{pmatrix}
        \hat{I} & 0\\
        0 & -\hat{I}
    \end{pmatrix}
\end{equation}
anticommutes with the Hamiltonian, i.e. $\hat{H}\hat{\Gamma}+\hat{\Gamma}\hat{H}=0$. 

This property ensures that each eigenstate $\ket{\psi}$ with eigenvalue $k_z$ is accompanied by another eigenstate $\hat{\Gamma}\ket{\psi}$ with the associated eigenvalue $-k_z$. As a consequence, the spectrum is symmetric with respect to the zero level defined by the propagation constant of $s$ or $p$ mode in an isolated waveguide, $k_0$. This feature of the spectrum is clearly seen in Fig.~2 of the main text.

In analogy to the well-celebrated Su-Schrieffer-Heeger model, chiral symmetry also guarantees the robustness against the perturbations that do not break this symmetry. 

\section{II. Pseudospin basis, Zak phases and winding numbers}

The designed model of interacting $s$ and $p$ photonic molecules can be interpreted in terms of effective pseudospin by constructing linear superpositions of orbitals $s+p$ and $s-p$. This interpretation is especially clear under two simplifying assumptions: (i) the propagation constants of $s$ and $p$ modes are equal, i.e. the modes are degenerate; (ii) $\kap_{ss}=\kap_{pp}$ which ensures an additional symmetry between $s$ and $p$ orbitals.

Given the two assumptions above, we switch to the hybrid basis via unitary transformation
\begin{equation}
    U=\frac{1}{\sqrt{2}}\,
    \begin{pmatrix}
        1 & 1 & 0 & 0\\
        0 & 0 & 1 & -1\\
        1 & -1 & 0 & 0\\
        0 & 0 & 1 & 1
    \end{pmatrix}\:.
\end{equation}
This transformation brings the Hamiltonian [Eq.~(1) of the main text] to the form
\begin{equation*}
    \hat{H}=
    \begin{pmatrix}
        \hat{H}_{+} & 0\\
        0 & \hat{H}_{-}
    \end{pmatrix}\:,
\end{equation*}
where the respective pseudospins are described by the Hamiltonian resembling canonical Su-Schrieffer-Heeger model with intra-cell couplings $\kap_{ss,2}\pm\kap_{sp,2}$ and inter-cell couplings $\kap_{ss,1}\mp\kap_{sp,1}$:
\begin{equation}
    \hat{H}_{\pm}=-
    \begin{pmatrix}
      0 & e^{-ik_x}\,(\kap_{ss,1}\mp\kap_{sp,1})+(\kap_{ss,2}\pm\kap_{sp,2})\\
      e^{ik_x}\,(\kap_{ss,1}\mp\kap_{sp,1})+(\kap_{ss,2}\pm\kap_{sp,2}) & 0
    \end{pmatrix}\:.
\end{equation}
Each of these spin blocks has a well-defined winding number and associated Zak phase. The overall invariant is recovered by adding up the contributions from the individual blocks. Studying the lattice with the integer number of unit cells, we distinguish three topologically distinct scenarios:

\noindent (1) For a geometrically dimerized lattice with smaller intra-cell distance $dx_1 < dx_2$ the couplings $\kap_{ss,1}$ and $\kap_{sp,1}$ that correspond to the distance $dx_2$ are smaller than $\kap_{ss,2}$ and $\kap_{sp,2}$, respectively [see Fig.~1(d) of the main text]. However, if 
\begin{equation}\label{eq:TransEq1}
    \kap_{ss,1}+\kap_{sp,1}>\kap_{ss,2}-\kap_{sp,2}\:,
\end{equation}  
one of the SSH submodels becomes nontrivial indicating the onset of nontrivial Zak phase and the emergence of a pair of edge-localized modes at two edges of the array. In terms of phase diagram [Fig.~2(d) of the main text, see the line $\lambda=730$~nm for perfect mode degeneracy], this means that the topological edge states may appear even for $dx_1<dx_2$ provided the condition Eq.~\eqref{eq:TransEq1} is met. On the other hand, if Eq.~\eqref{eq:TransEq1} is not fulfilled, the edge states are absent.

\noindent (2) In the case of a homogeneous lattice with $dx_1=dx_2$ when $\kap_{ss,1} = \kap_{ss,2}, \, \kap_{sp,1} = \kap_{sp,2}$, the dimerizations of the two pseudospin submodels are opposite, so for any finite lattice one of the pseudospin blocks has nontrivial Zak phase, and there is a single edge state at each of the lattice edges. 

\noindent (3) Finally, if the dimerization of the lattice is $dx_1 > dx_2$, the couplings $\kap_{ss,1}$ and $\kap_{sp,1}$ corresponding to $dx_2$ distance are larger. However, if 
\begin{equation}\label{eq:TransEq2}
\kap_{ss,1} - \kap_{sp,1}<\kap_{ss,2} + \kap_{sp,2}\:,  
\end{equation}
only one of the pseudospin blocks possesses a nontrivial Zak phase and hence there is still a single pair of edge-localized modes. In terms of phase diagram [Fig.~2(d) of the main text, see the line $\lambda=730$~nm for perfect mode degeneracy], this means that there is a single pair of edge-localized states even for $dx_1>dx_2$ provided the condition Eq.~\eqref{eq:TransEq2} is met. If Eq.~\eqref{eq:TransEq1} is not fulfilled, the designed array hosts {\it two pairs} of edge-localized modes, since both pseudospin blocks possess nontrivial Zak phase.

The three regimes described above qualitatively correspond to three topologically distinct regions with $P=0, \, 0.5$ and $1$ depicted in Fig.~2(c) in the main text. 

Next, we generalize our treatment to the experimentally relevant case when $\kap_{ss}$ and $\kap_{pp}$ couplings are not equal, while $s$ and $p$ modes are still assumed degenerate: $\delta k_z=0$. In such case, the Hamiltonian does not decouple into two independent pseudospins. However, it can be brought to the off-diagonal form [Eq.~(1) of the main text with $\hat{\beta} = 0$~\cite{Ryu2010Jun,Chiu2016Aug}], which allows us to compute the so-called winding number.

\begin{figure*}[h!]
	\centering
	\includegraphics[width=0.70\textwidth]{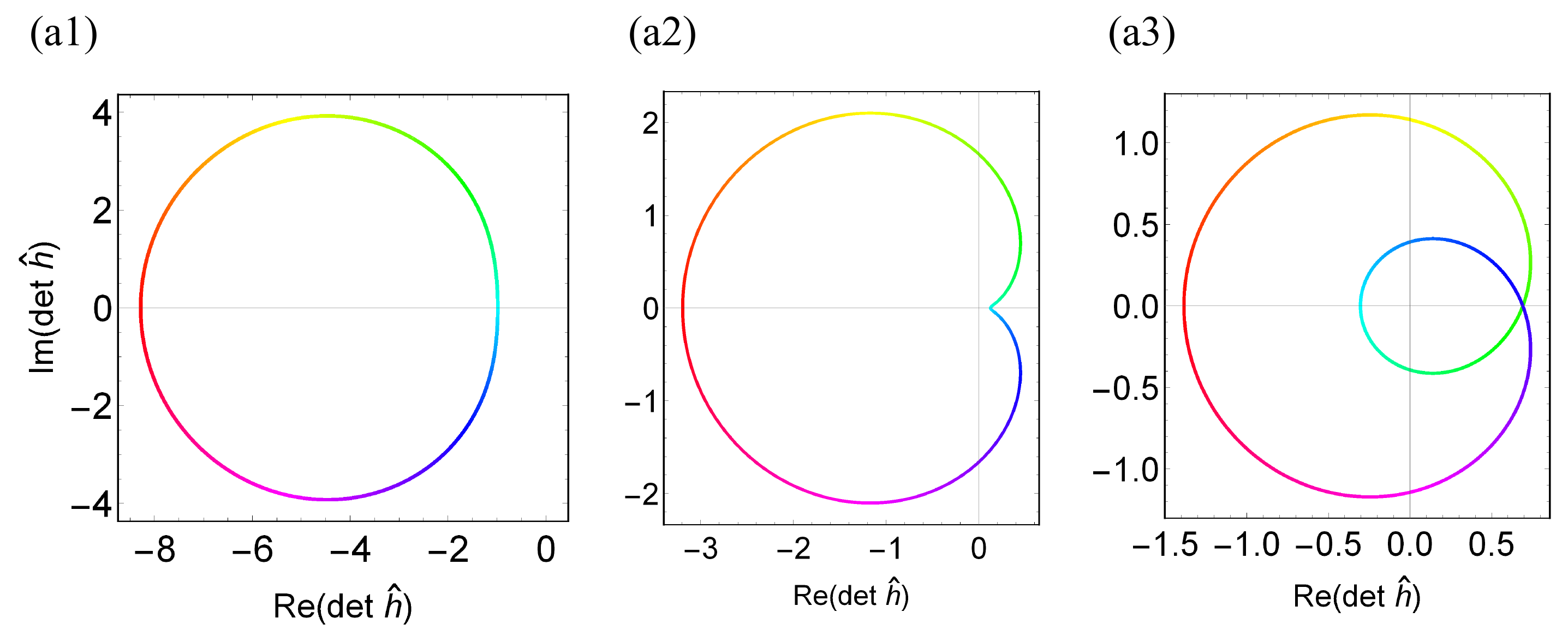}
	\caption{ 
        Paths of the determinant of the off-diagonal part $\det{\hat{h}}$ of the Hamiltonian (Eq.(1) in the main text) on the complex plane, demonstrating winding number calculation, at perfect mode degeneracy $\delta k_{z0}$ for 
        $dx_1=19$~$\mu$m (a1) [$W=0$], $25$~$\mu$m (a2) [$W=1$] and $33$~$\mu$m (a3) [$W=2$] (other parameters correspond to Fig.~2 in the main text). Color encodes different values of $k_x$ swept from $-\pi$ to $\pi$. 
        }
	\label{fig:S2}
\end{figure*}

 Figure~\ref{fig:S2} shows the evolution of the determinant of the upper off-diagonal block $\hat{h} = e^{-i k_x} \hat{\kappa}_{1-} + \hat{\kappa}_{2+}$ on the complex plane as $k_x$ sweeps through the 1D Brilloin zone in three topologically distinct cases, with winding number equal to number of times $\det{\hat{h}}$ encircles the origin. These results confirm the qualitative reasoning above, capturing three topologically distinct phases. In turn, full topological phase diagram can be further reconstructed by tracking band closing and reopening in the $(\lambda, \, dx_1)$ parameter space. 

\section{III. Wannier center polarization}

Another, and most general approach to calculate the topological invariant is based on summing up the Wilson loop eigenvalues for the two lowest occupied bands, thereby extracting Wannier center polarization~\cite{Benalcazar_2017_PRB,Benalcazar_2017_Science}. We calculate the Wilson loop matrix as $ \hat{W}  = \hat{F}_{k_{x1}} \hat{F}_{k_{x2}} ... \hat{F}_{k_{xN}} $, where elements 
$\hat{F}_{k_{xN}} = \hat{U} \hat{V}^\dag$ are constructed from the singular-value decomposition $ \hat{G} = \hat{U} \hat{D} \hat{V}^\dag $ of the Wilson line elements $\left[G_{k_{xj}} \right]^{m n}=\left\langle u_{k_{xj} + \Delta_k}^n \mid u_{k_{xj}}^m\right\rangle$, where upper indices mark the occupied bands, and $\Delta k = 2 \pi / N$. Here, $N$ denotes the number of $k$-points in the Wilson loop, while $\ket{u_k}$ stands for the periodic part of the Bloch function.

Eigenvalues of the Wilson loop matrix $\hat{W}$ have form $e^{i 2\pi \nu_{1,2}}$, where $\nu_{1,2}$ mark the positions of Wannier centers~\cite{Benalcazar_2017_PRB}. The total Wannier center polarization is then the the sum of the Wannier centers, $P  = \nu_{1} + \nu_{2}$. The results of this calculation are shown in Fig.~2(c) in the main text and are fully consistent with the two approaches outlined above.

\section{IV. Evolution of finite lattice spectrum with and without disorder}

In this section, we present a set of plots describing the evolution of finite lattice spectra with and without disorder to support and complement the local density of states plots in the main text and exemplify the robustness of the edge states against disorder.

\begin{figure*}[h]
	\centering
	\includegraphics[width=1.0\textwidth]{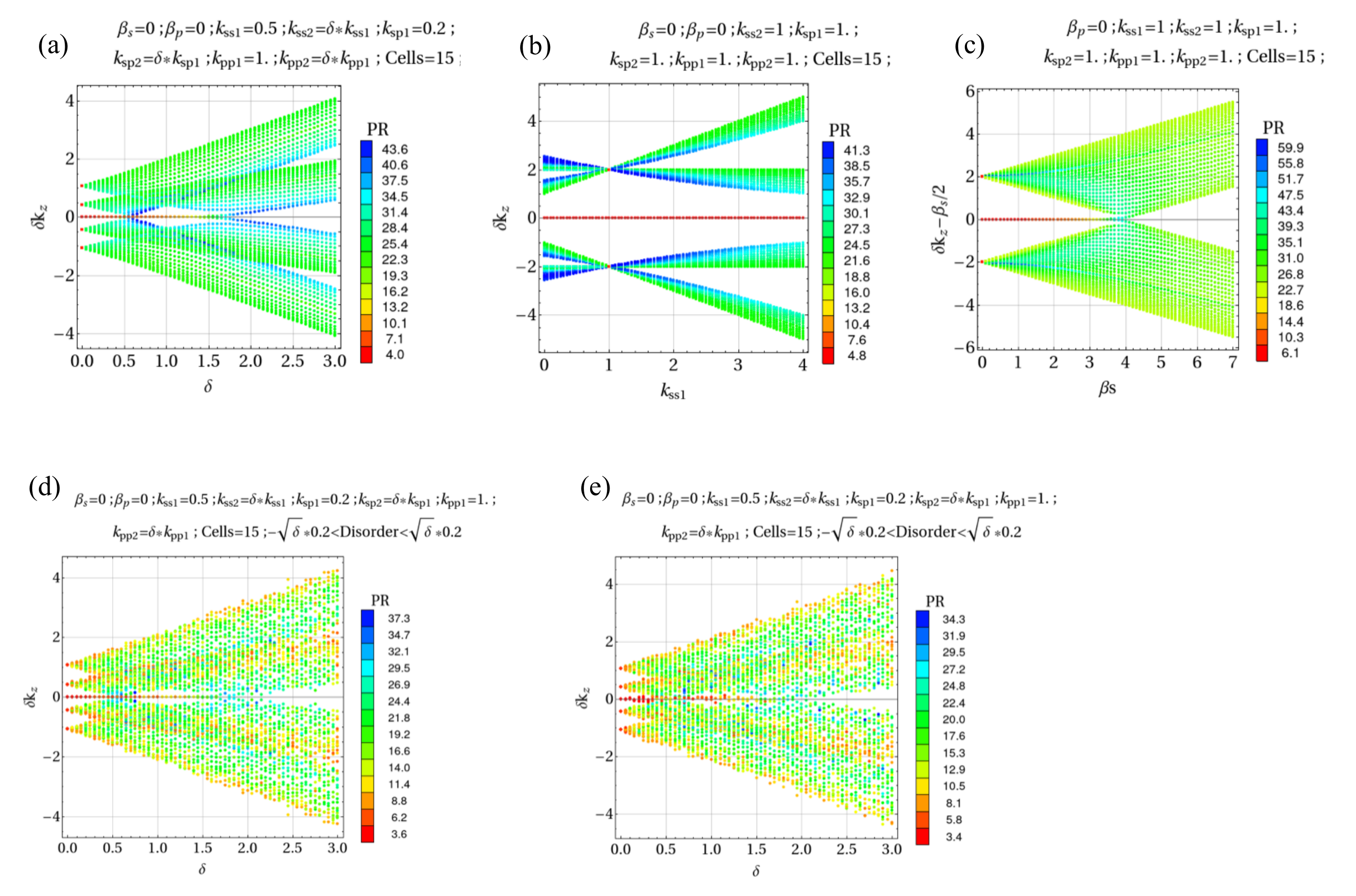}
	\caption{(a-c) Evolution of finite-lattice spectra for varying couplings~(a-b) and propagation constants~(c) (see details in the text of the Section). (d-e) Similar evolutions for varying couplings, but with two types of disorder included: disorder respecting teh chiral symmetry $\hat{\Gamma}$~(d) and disorder in propagation constants~(e). Color marks the participation ratio (PR) with red corresponding to the edge-localized states.}
	\label{SMEvolutions}
\end{figure*}

Fig.~\ref{SMEvolutions} shows several spectral evolution scenarios:  

(a) Perfect mode degeneracy, with coupling parameters varied 
proportionally, parameterized by one corresponding control parameter $\delta$. Two topological transitions are clearly identified. From $\delta=0$ (corresponding to detached edges) there are $4$ edge states; as $\delta$ is after $\delta=0.5$, only $2$ edge states remain in the bandgap, and as $\delta$ is increased further after $\delta=1.7$, the edge states disappear, indicating a topologically trivial phase.

(b) Perfect mode degeneracy, with only $\kappa_{ss1}$ coupling parameter varied. There are four flat bands when $\kappa_{ss1}$ when all other couplings are equal to $1$, along with two edge states (corresponding to the first topological phase with $P=0.5$). Therefore, for fine-tuned coupling constants topological flat bands appear, absent in the single-mode SSH model.

(c) Varying mode propagation constants. We observe a topological transition from the first topological to the trivial phase as we move further from the perfect mode degeneracy.

(d) An eigenvalue evolution plot with hermitian chiral-symmetry-preserving disorder in the couplings, corresponding to the disorder in the distances between photonic molecules. We observe two transitions with the edge states at the same zero energy as in~(a). 

(e) Another eigenvalue evolution plot with disorder in the propagation constants (or, equivalently, disorder in refractive index contrast). As this type of disorder does not preserve chiral symmetry, the edge states appear to be more volatile and hybridize with the bulk bands near the topological transition points.

To sum up, the numerical calculations confirm the existence of the three topological phases, the possibility to tune into different topological regimes via geometric dimerization and wavelength, existence of topological all-flat bands for fine-tuned couplings, and resilience of the edge states to the experimentally relevant types of disorder, especially to ones preserving the essential symmetries of the model, such as the chiral symmetry which quantizes the invariant,~$P$.

\section{V. Coupling constants characterization}


To calculate the couplings versus inter-waveguide horizontal distance $dx$ and operating wavelength $\lambda$, we calculate longitudinal wavenumbers $\delta k_z$ of dimer eigenstates for the relevant geometries (as the dimers appear in the lattice) in COMSOL Multiphysics. 

\begin{figure*}[h]
	\centering
	\includegraphics[width=1.0\textwidth]{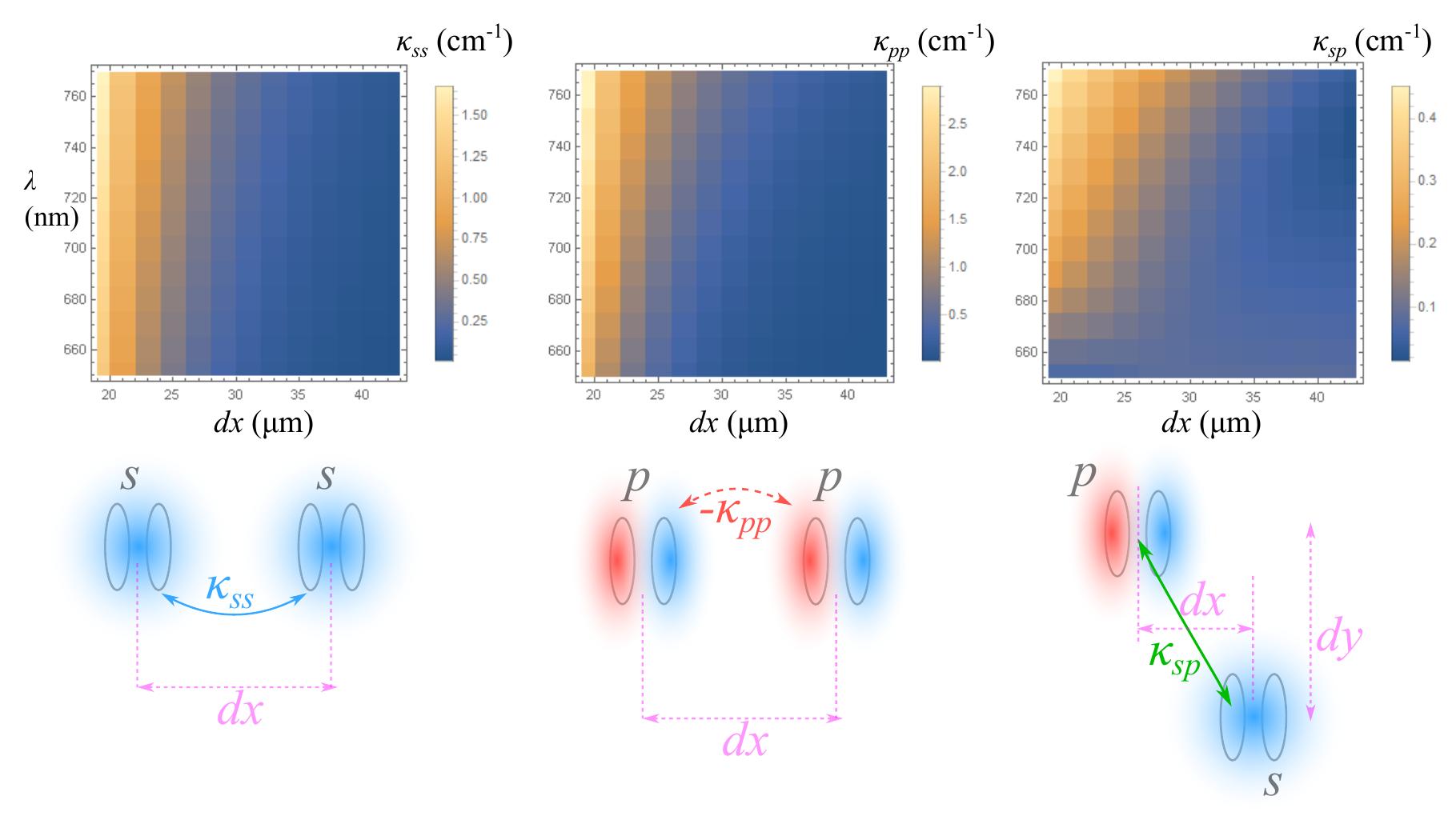}
	\caption{Couplings extracted from dimer longitudinal wavenumbers calculated numerically in COMSOL Multiphysics ($dy = 20 \,\mu$m as in experiment for the $\kappa_{sp}$ calculation; other parameters similar to Fig.2 in the main text).}
	\label{SMcouplings}
\end{figure*}

The eigenvalues are then matched to the tight-binding model which accounts for the imperfect mode degeneracy $\hat{\beta}$. Specifically, the eigenmodes for the $s$-$s$ and $p$-$p$ horizontal dimers read $\delta k_{z\,ss,pp} = \pm \kappa_{ss,pp}$, while $\delta k_{z\,sp} = \pm \sqrt{ \kappa_{sp}^2 + \delta \beta^2(\lambda)/4 }$ where $\delta \beta$ is the difference of the longitudinal wavenumbers for $s$ and $p$ modes (with additional polarization degeneracy in all three cases). 
The result is shown in Fig.~\ref{SMcouplings}. We use the obtained couplings to obtain the topological band diagram in Fig.2(c), as well as numerical edge intensity in Fig.3(c) in the main text.

\section{VI. SP femtosecond photonic lattices}

\begin{figure*}[h]
	\centering
	\includegraphics[width=0.5\textwidth]{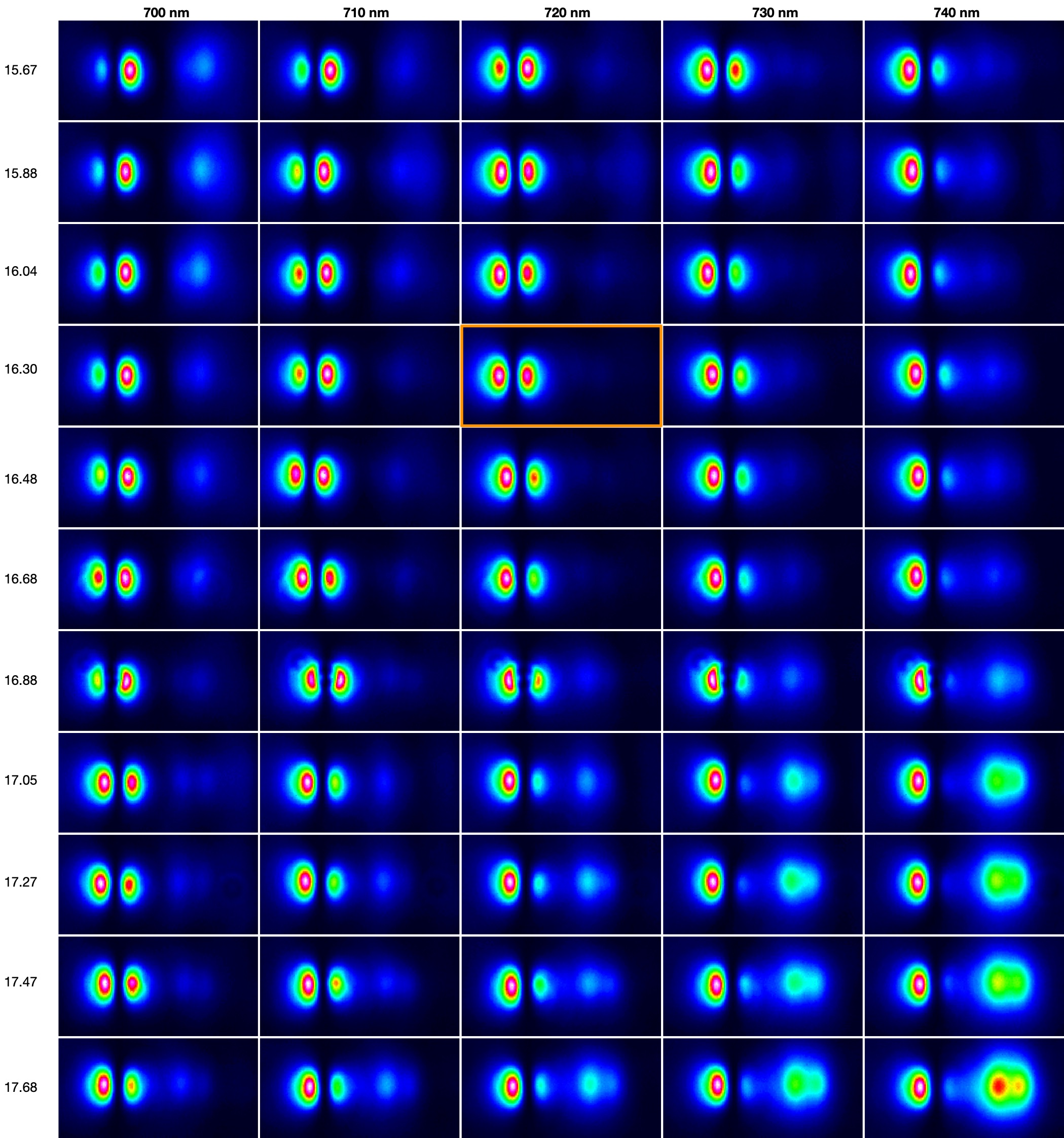}
	\caption{$sp$ dimers calibration.}
	\label{SMcal}
\end{figure*}

By using the femtosecond laser writting technique~\cite{Davis96,Szameit2005,Flamini2015} with a $1030$ nm infrared laser, having an average pulse width of $215$ fs and a repetition rate of $0.5$ MHz, we first fabricate several $sp$ couplers. We look for optimized fabrication parameters such that we can observe a strong coupling between $s$ and $p$ molecular modes on a horizontal configuration. After several experiments, we found a good regime with $s$ molecules composed of two close waveguides, with a center to center distance of $7\ \mu$m and an average power of $\sim 100$ mW, while the $p$ molecules are obtained for an average power of $\sim 170$ mW. All waveguides in this work were fabricated at a writing velocity of $0.9$ mm/s. All the parameters were optimized such that a zero detuning condition ($\delta k_{z0} = 0$) was achieved around a wavelength of $730$ nm, which is an optimal parameter in our fabrication scheme in terms of light guidance. An example of this calibration is shown in Fig.~\ref{SMcal}. This figure shows a compilation with several $sp$ dimers having a full length $s$ molecule (right) and a $12$ mm long $p$ molecule (left). The experiments consist on exciting the $s$ molecule while varying the input wavelength (at around $730$ nm) and look for the optimal coupling efficiency in between both molecules. At the left column of this figure we show a value which is proportional to the fabrication power for the $s$ molecule, while keeping the $p$ molecule fabrication power as constant. We found a good fabrication regime for a value at around $16$, where most of the energy has been efficiently transferred from the $s$ to the $p$ photonic molecule.

\begin{figure*}[]
	\centering
	\includegraphics[width=0.5\textwidth]{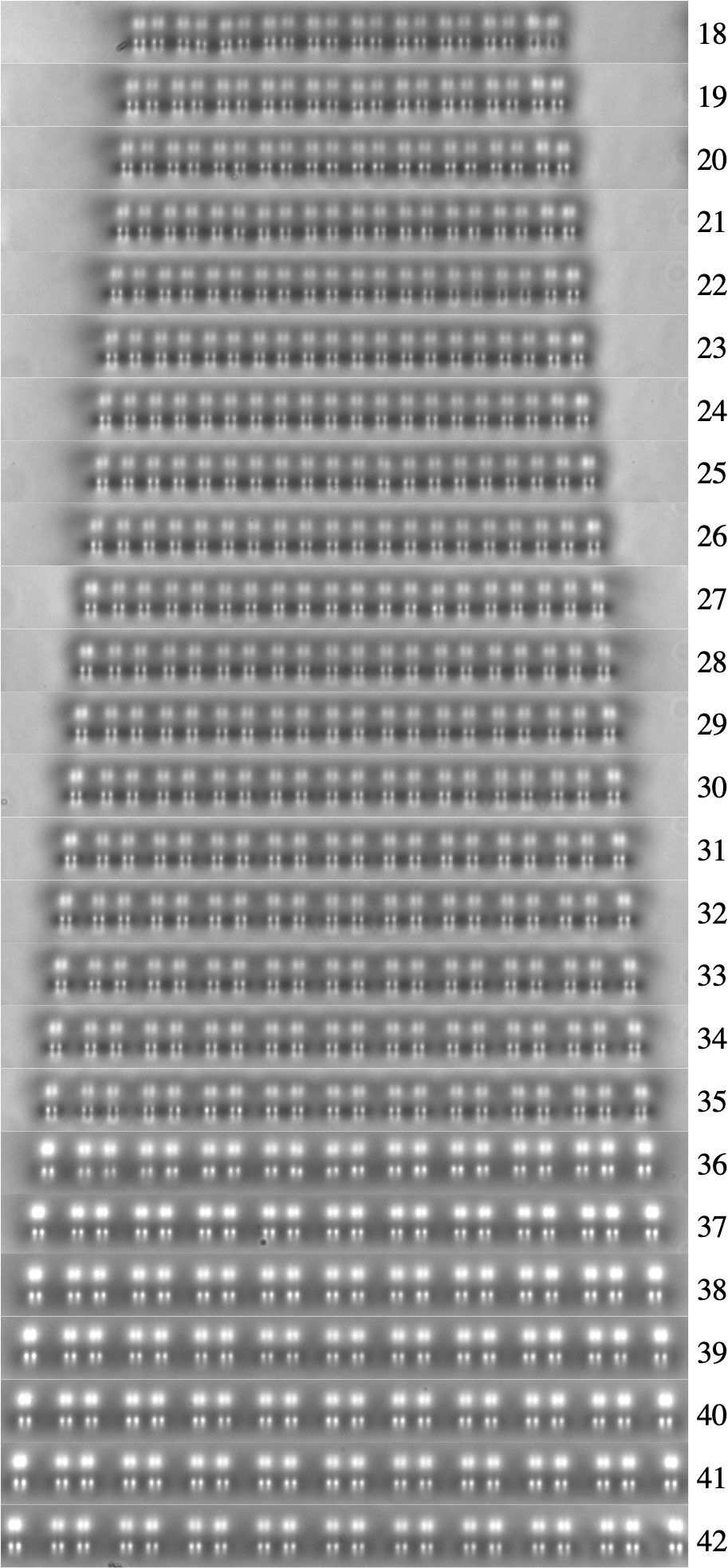}
	\caption{Microscope images, after white light illumination, of the femtosecond fabricated lattices with the distance $dx_1\ [\mu m]$ indicated to the right.}
	\label{fig:SMlatt}
\end{figure*}

Once we have set the optimal fabrication parameters for having a strong interaction in between the $s$ and the $p$ molecules around $730$ nm, we fabricated several SP molecular lattices having 10 dimerized unit cells. As every photonic molecule is formed by two close waveguides, every $sp$ lattice with 40 lattice sites is composed of 80 waveguides in total. Fig.~\ref{fig:SMlatt} shows the final 25 fabricated lattices with parameters $dx_1\in\{18,42\}\ \mu$m, $dx_2=25\ \mu$m and $dy=20\ \mu$m, on a $70$ mm-long glass wafer. It can be noticed in this figure how every $s$ (top files) and $p$ (bottom files) photonic molecules are composed of two close waveguides, in this case at a distance of $7\ \mu$m. In particular, we can observe a more complex structure for $p$ molecules due to the larger fabrication power. As a white light illumination contains a full set of visible wavelengths, then the photonic structure is able to support even higher order orbital states for shorter wavelengths~\cite{Silva_PRL_2021,Caceres-Aravena2022Jun}.

\begin{figure*}[]
	\centering
	\includegraphics[width=0.8\textwidth]{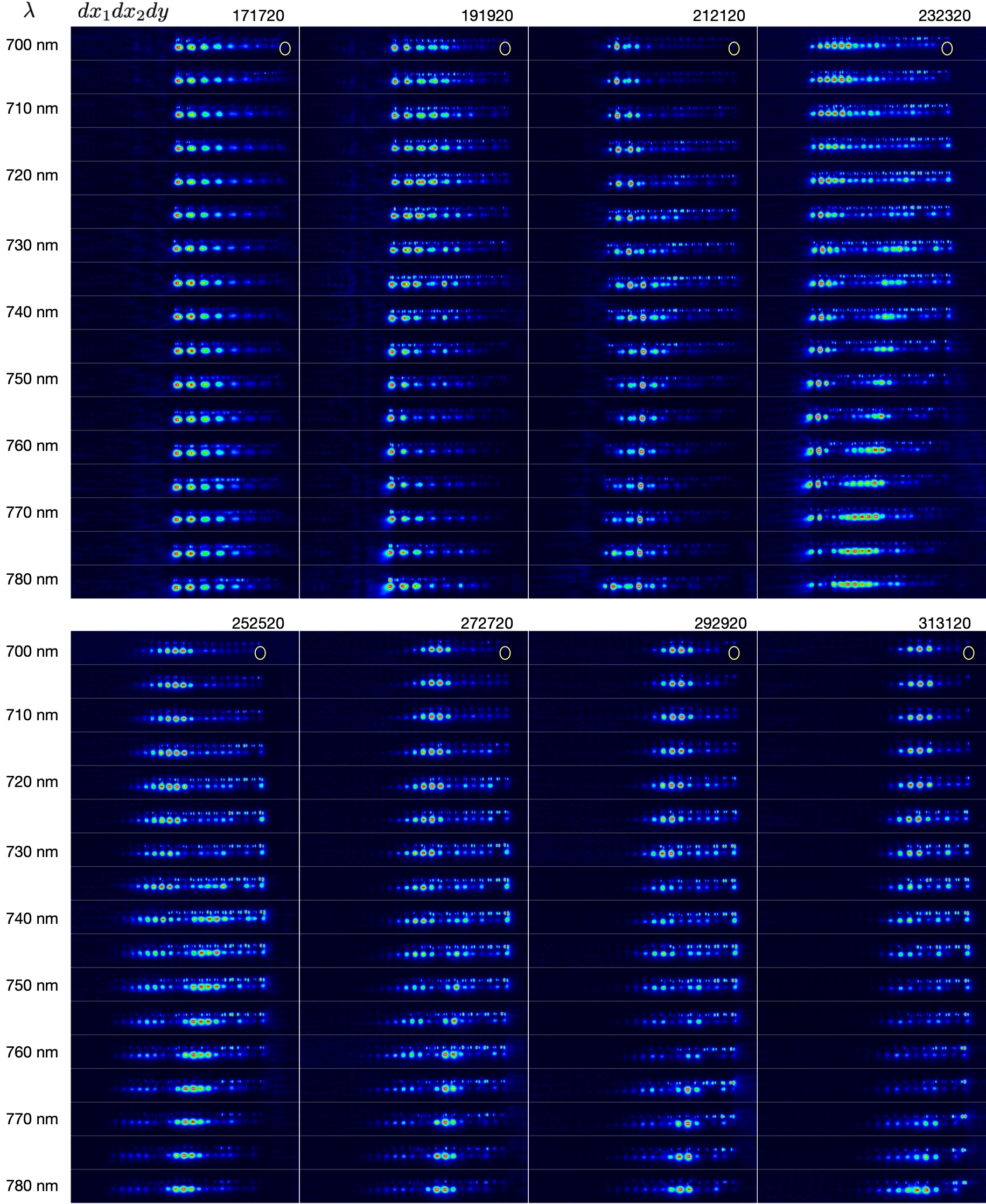}
	\caption{Output intensity profiles after a $s$ molecule edge excitation of 8 homogeneous lattices, having a different $dx_1=dx_2$ distance and a fixed $dy=20\ \mu$m, as indicated at the top-right of every vertical panel. The yellow ellipses indicate the input position. Every lattice was excited with a set of wavelengths in the interval $\lambda\in\{700,780\}$ nm, as indicated to the left.}
	\label{fig:homos}
\end{figure*}

We optimized the lattice dimensions after performing several fabrication runs, such that the light propagation becomes clear and simple for the number of lattice sites and glass length. We can see in Fig.~\ref{fig:homos} that for a shorter lattice spacing ($dx_1=dx_2<25\ \mu$m), the light already arrives to the opposite edge (left edge, in this case) and the reflections contaminate the output intensity profiles. We notice that for a lattice with $dx_1=dx_2=25\ \mu$m, the light spreads smoothly through the system without any noticeable reflection feature. For larger spacing lattices ($dx_1=dx_2>25\ \mu$m), we observe that the light does not explore the whole system and show narrower diffraction patterns. Therefore, as it becomes evident in this figure, for the number of lattice sites chosen ($40$) in our experiment and the available glass length ($70$ mm), the dimensions $dx_1=dx_2=25\ \mu$m were the optimal ones.

\bibliography{refs}